\documentclass[aps,prb,showpacs,twocolumn,showpacs]{revtex4-1}

\usepackage{graphicx}
\usepackage{amsmath}
\usepackage{amssymb}
\usepackage{color}

\definecolor{lila}{rgb}{0.5,0,1}
\newcommand{\bnen}{\begin{equation}}
\newcommand{\eden}{\end{equation}}
\newcommand{\bean}{\begin{eqnarray}}
\newcommand{\eean}{\end{eqnarray}}

\newcommand{\bna}{\begin{array}}
\newcommand{\eda}{\end{array}}

\begin{document}

\title{Interorbital interaction in the one-dimensional periodic Anderson
model:\\ A density-matrix renormalization-group study}

\author{I. Hagym\'asi$^{1,2}$}
\author{J. S\'olyom$^{1}$}
\author{\"O. Legeza$^1$}

\affiliation{$^1$Strongly Correlated Systems "Lend\"ulet" Research Group, Institute for Solid State
Physics and Optics, MTA Wigner Research Centre for Physics, Budapest H-1525 P.O. Box 49, Hungary\\
$^2$Department of Theoretical Physics, University of Szeged, Tisza Lajos krt 84-86, H-6720 Szeged,
Hungary}

\date{\today}

\begin{abstract}
We investigate the effect of the Coulomb interaction, $U_{cf}$, between the conduction and f
electrons in the periodic Anderson model using the density-matrix
renormalization-group algorithm.  We
calculate the excitation spectrum of the half-filled symmetric model with an emphasis on
the spin
and charge excitations. In the one-dimensional version of the model it is found that the spin
gap is
smaller than the charge gap below a certain value of $U_{cf}$ and
the reversed inequality is valid for stronger $U_{cf}$. 
This behavior is also verified by the
behavior of the spin and density correlation functions. We also
perform a quantum
information analysis of the model and determine the entanglement map of the f
and conduction
electrons. It is
revealed that for a certain $U_{cf}$ the ground state is dominated by the configuration in which
the conduction and f electrons are strongly entangled, and the ground state is almost a product
state. For larger $U_{cf}$ the sites are occupied alternatingly dominantly by two f electrons or by
two conduction electrons.

\end{abstract}


\maketitle

\section{Introduction} Kondo insulators are a peculiar group of rare-earth
materials, which behave as metals with magnetic moments above a characteristic temperature ($\sim
100$ K) and become semiconductors at low temperatures due to strong
correlations.\cite{Kondo_ins_review} Gaps are opened in
both the spin and charge sectors, and they are in the order
of a few meV, which defines the small energy scale of these compounds. These small
gaps cannot be
understood in a simple band picture, since the strong interaction between the
electrons plays a
crucial role. Furthermore this is the reason why different energy scales arise
in the spin and
charge sectors.
\par The minimal model for the Kondo insulators is the half-filled periodic Anderson
model\cite{Review} (PAM). In its one-dimensional version the
Hamiltonian reads
\begin{equation}   \begin{split}
     \mathcal{H}_{\rm PAM} = & -t\sum_{j,\sigma}
       (\hat{c}_{j\sigma}^{\dagger}
	   \hat{c}^{\phantom \dagger}_{j+1\sigma}+\hat{c}_{j+1\sigma}^{\dagger}
	   \hat{c}^{\phantom \dagger}_{j\sigma})\\
&-V\sum_{j,\sigma}(\hat{f}_{j\sigma}^{\dagger}
	  \hat{c}^{\phantom \dagger}_{j\sigma}
    +\hat{c}_{j\sigma}^{\dagger} \hat{f}^{\phantom \dagger}_{j\sigma}) 
+\varepsilon_f\sum_{j,\sigma}\hat{n}^f_{j\sigma}\\
&+U_f\sum_{j}\hat{n}^f_{j\uparrow}
\hat{n}^f_{j\downarrow}
\label{PAM:Hamiltonian}
	     , 
\end{split}
\end{equation}
where the notation is standard and $W=4t$ is taken as the energy unit.
The spin and charge gaps of the one-dimensional PAM have been studied by several methods in the
past decades.  
The exact diagonalization studies\cite{Ueada:exact} pointed out that both the
charge and spin gaps are finite.  This analysis was later confirmed
and refined by the
density-matrix renormalization-group calculations
\cite{Guerrero:DMRG,KLM:DMRG,Santos:DMRG}. Both methods showed that the charge
gap is
always larger than the spin gap, and decreases much more slowly than the spin
gap as $U_f$ is
incresed and
their ratio diverges in the large $U_f$ limit. Later on, this inequality was rigorously proven for the ordinary
periodic Anderson model and it was shown that it remains valid for a $d$-dimensional simple cubic
lattice as well. \cite{Tian:proof}
The charge and spin gaps are also directly measurable quantities using optical
and neutron
scattering
measurements, respectively. It is worth noting that for numerous compounds
 the experimental ratio of
the gaps is less than one, which is in qualitatively good agreement with the theoretical
predictions for the one-dimensional PAM, and in some cases, like CeRu$_4$Sb$_{12}$ or
CeFe$_4$Sb$_{12}$, even quantitative agreement can be achieved with the theoretical
predictions.\cite{Takabatake:review}
However, for CeRhAs the ratio
of the spin\cite{CeRhAs:spingap} and charge gaps\cite{CeRhAs:chargegap} was
found to be greater than one, namely, $\Delta_s/\Delta_c\approx1.5$, which cannot be
understood at all in the frame of the ordinary periodic Anderson model.\cite{Takabatake:review} One
of the major aims of the present paper is to provide a possible explanation to the problem.
\par Several extensions of the PAM have been considered lately, to model the
effect of further electron-electron interactions. The role of conduction
electron interaction in the Kondo lattice model and PAM has been thoroughly
investigated.\cite{KLM:Ud,Schork:cikk}
Recently, the correlations between conduction and f electrons have been shown to play an
important role to understand the 
critical valence fluctuations.\cite{Miyake:review} This interaction term leads to the
extended PAM (EPAM):
\begin{equation}
 \mathcal{H} =\mathcal{H}_{\rm PAM}+U_{cf}\sum_{j,\sigma,\sigma'}
\hat{n}^f_{j\sigma}\hat{n}^c_{j\sigma'}.
\label{eq:EPAM}
\end{equation}
This model has been investigated by several modern techniques
recently,\cite{DMRG:cikkek,Hirashima:cikk,PBR:cikk,Kubo:GW,Kawakami:Udf_ins}
and the valence transition has been explained
successfully. However, less attention has been paid
to the
Kondo insulator
case.\cite{Kawakami:Udf_ins,Hagymasi:Udf} It has been shown
recently using dynamical mean-field theory (DMFT), which is exact for infinitely large dimensions
that in the symmetric case for small hybridization ($V\ll W$) the model displays antiferromagnetic
order for small $U_{cf}$ which, however, disappears for large $U_{cf}$ and charge order develops. In the charge 
ordered phase doubly occupied c and f sites appear in an alternating fashion, since the c and f electrons tend to avoid each other.
Since these
results were obtained via DMFT, which neglects spatial fluctuations, one can
naturally ask what happens in low-dimensional systems, where the fluctuations
are more important.
\par Our purpose in this paper is therefore to examine the one-dimensional EPAM. Naturally, we do
not
address the possibility of the presence of long-range order that was found in infinite dimensions.
We apply the density-matrix
renormalization-group method\cite{White:DMRG} (DMRG), which is a powerful tool to find the ground state and the first few excited states. 
Further advantage of
the DMRG method is that we can determine the von Neumann
entropies\cite{legeza2003b,vidallatorre03,calabrese04,luigi2008} of single and multisite subsystems, which
turned out to be very good
indicators of drastic changes in the
wavefunction.\cite{mfyang,legeza2006,laflorence,vidal} They are known to exhibit
anomalous behavior where the character of the ground state
changes dramatically.
\par The setup of the paper is as follows. In Sec. II. our numerical results are
presented for the spin and the charge gaps of the extended periodic Anderson model. In
Sec. III. A
 we investigate the
spin and
density correlation functions. In Sec. III. B we perform a quantum information
analysis and determine the
entanglement map of the EPAM using the mutual information\cite{rissler2006,barcza2010,Boguslawski}
to get a physical picture for the ground state. Lastly, in Sec. IV. our conclusions are
presented.

\section{Spin and charge gaps}
\begin{figure}[!htb]
\includegraphics[scale=1]{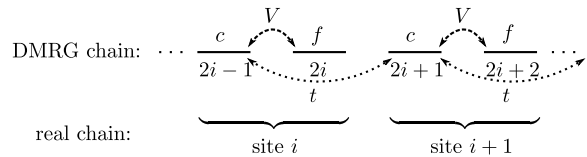}
\caption{Sketch of the DMRG implementation of the model Hamiltonian Eq. (\ref{eq:EPAM}).}
\label{fig:model_abra}
\end{figure}

For convenience the EPAM has been implemented in
the DMRG procedure as a
generalized
Hubbard model with a special topology. The site $i$ with both c and f
electrons is replaced by two DMRG sites, one for the conduction electrons, the other for the f
electrons, and therefore instead of working with 16 states per site, only 4 states per DMRG sites
have to be considered. These $|\alpha\rangle$ states are the empty, singly occupied with down and
up spin, and the doubly occupied states, denoted by $|0\rangle$, $|\downarrow\rangle$,
$|\uparrow\rangle$, $|\uparrow\downarrow\rangle$, respectively, 
\begin{align}
\hspace{-0.38cm}
\begin{aligned}
|1\rangle^{(c)}=&|0\rangle,\\
|2\rangle^{(c)}=&|\downarrow\rangle^{(c)}_i=\hat{c}_{i\downarrow}^{\dagger}|0\rangle,\\
|3\rangle^{(c)}=&|\uparrow\rangle^{(c)}_i=\hat{c}_{i\uparrow}^{\dagger}|0\rangle, \\
|4\rangle^{(c)}=&|\uparrow\downarrow\rangle^{(c)}_i=\hat{c}_{i\uparrow}^{\dagger}\hat{c}_{
i\downarrow }^{\dagger}|0\rangle,
\end{aligned}
&
\begin{aligned}
|1\rangle^{(f)}=&|0\rangle, \\
|2\rangle^{(f)}=&|\downarrow\rangle^{(f)}_i=\hat{f}_{i\downarrow}^{\dagger}|0\rangle,\\
|3\rangle^{(f)}=&|\uparrow\rangle^{(f)}_i=\hat{f}_{i\uparrow}^{\dagger}|0\rangle,\\
|4\rangle^{(f)}=&|\uparrow\downarrow\rangle^{(f)}_i=\hat{f}_{i\uparrow}^{\dagger}\hat{f}_{
i\downarrow }^{\dagger}|0\rangle.
\end{aligned}
\label{eq:states}
\end{align}
The schematic structure is shown in
Fig. \ref{fig:model_abra}. 
In what follows the chain length $N$ is understood as the
number
of
real EPAM sites. 
In our DMRG calculations,  we apply the dynamic block-state selection
algorithm\cite{DBSS:cikk1,DBSS:cikk2} and keep block
states up to 2000, the typical truncation errors are $10^{-6}-10^{-8}$. Open
boundary
condition is applied, and we considered chains up to a maximum length $N=50$ and finite-size scaling
is used to extrapolate the
quantites to the thermodynamic limit.
\par
\begin{figure}[!htb]
\includegraphics[scale=0.5]{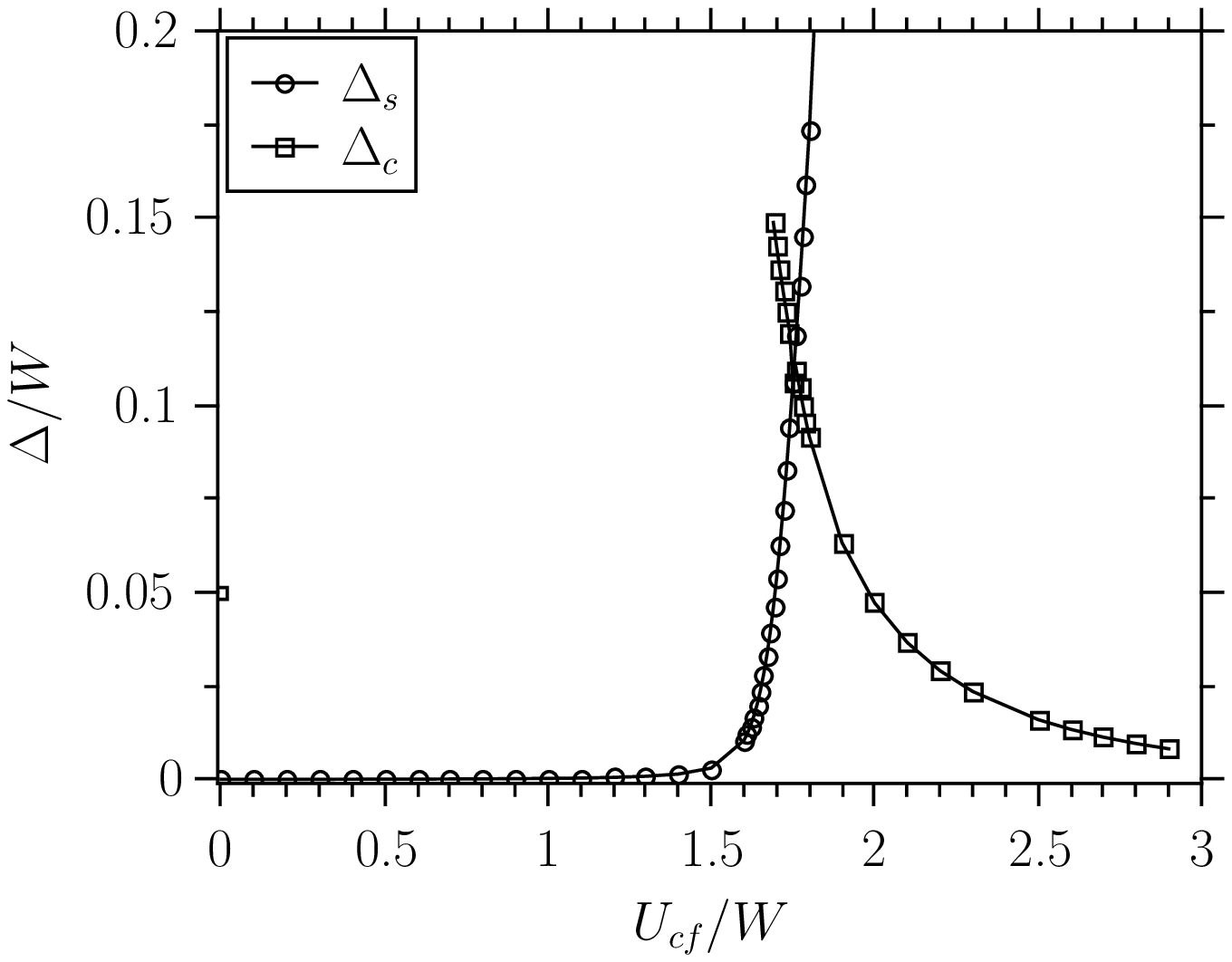}\\
\includegraphics[scale=0.5]{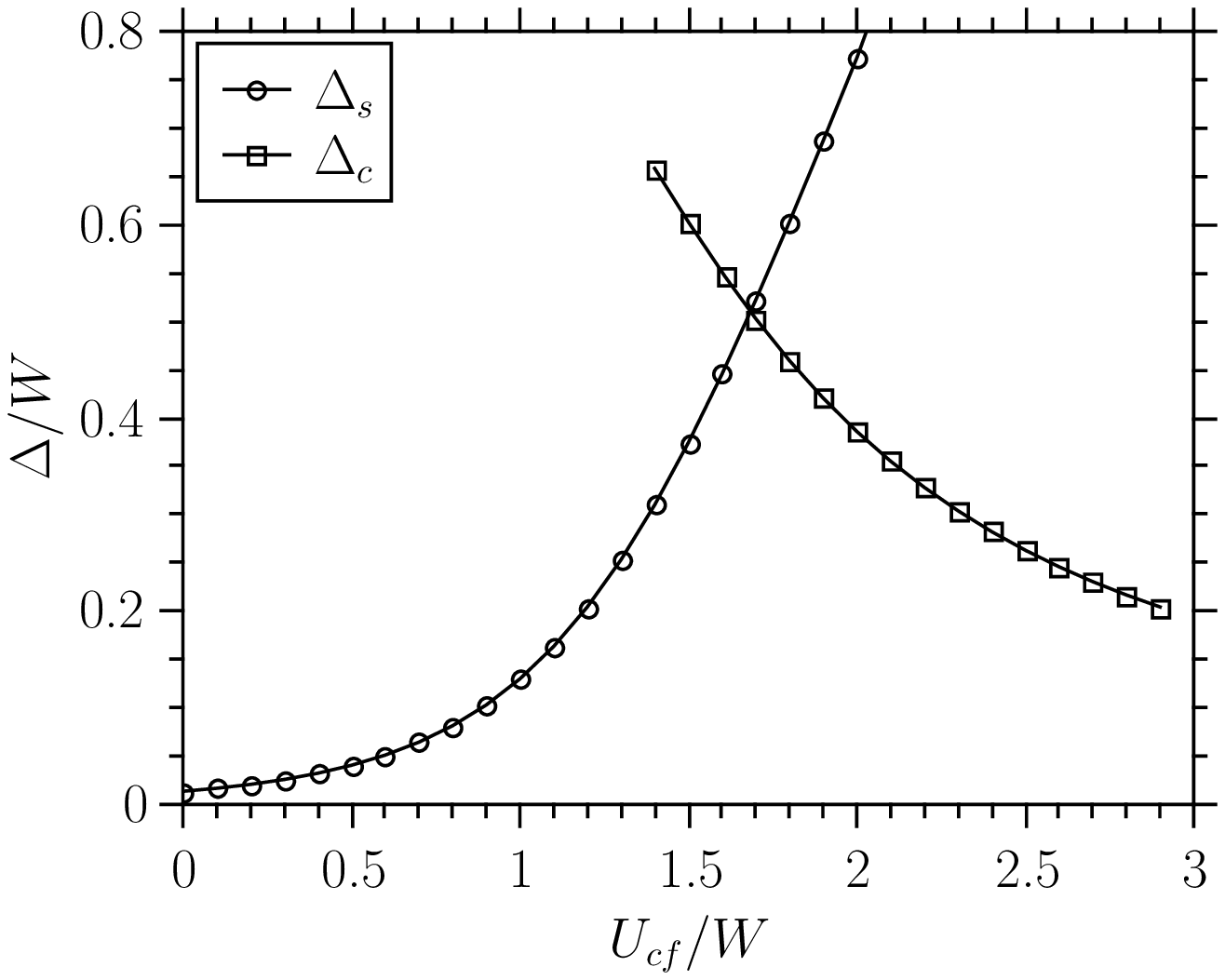}
\caption{The spin and charge gaps extrapolated to the thermodynamic limit as a function of
$U_{cf}$ for $V/W=0.1$ (upper panel), $V/W=0.3$ (lower panel) and
$U_f/W=3$.  The
lines are
guides to the eye.}
\label{fig:gaps}
\end{figure}

\begin{figure}[!htb]
\includegraphics[scale=0.55]{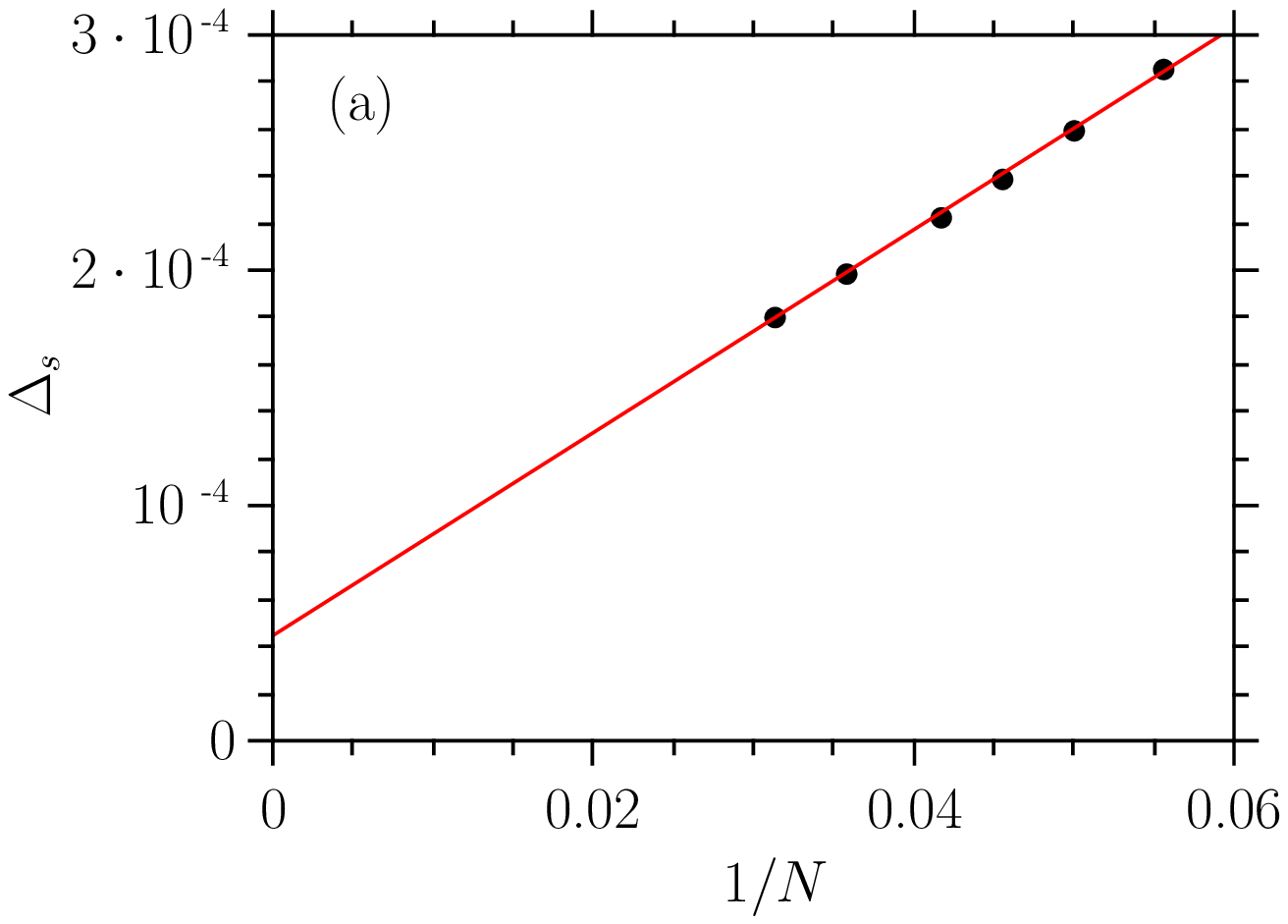}\\
\includegraphics[scale=0.55]{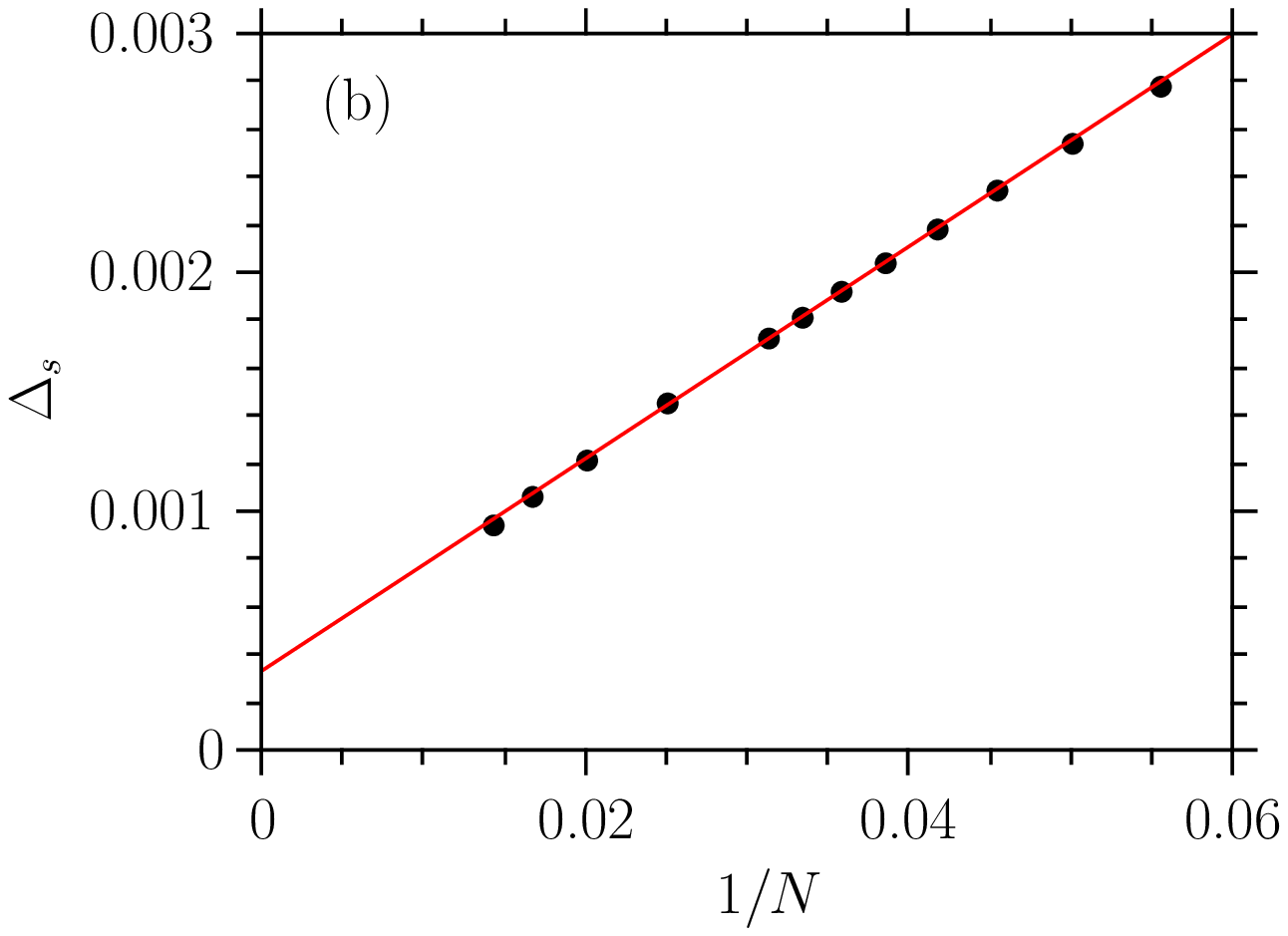}

\caption{Finite-size extrapolation of the spin gap for $U_{cf}/W=1$ (panel (a)), $U_{cf}/W=1.5$
(panel (b)). The solid lines denote the linear fits to the data.}
\label{fig:gap_scaling}
\end{figure}

\begin{figure}[!htb]
\includegraphics[scale=0.5]{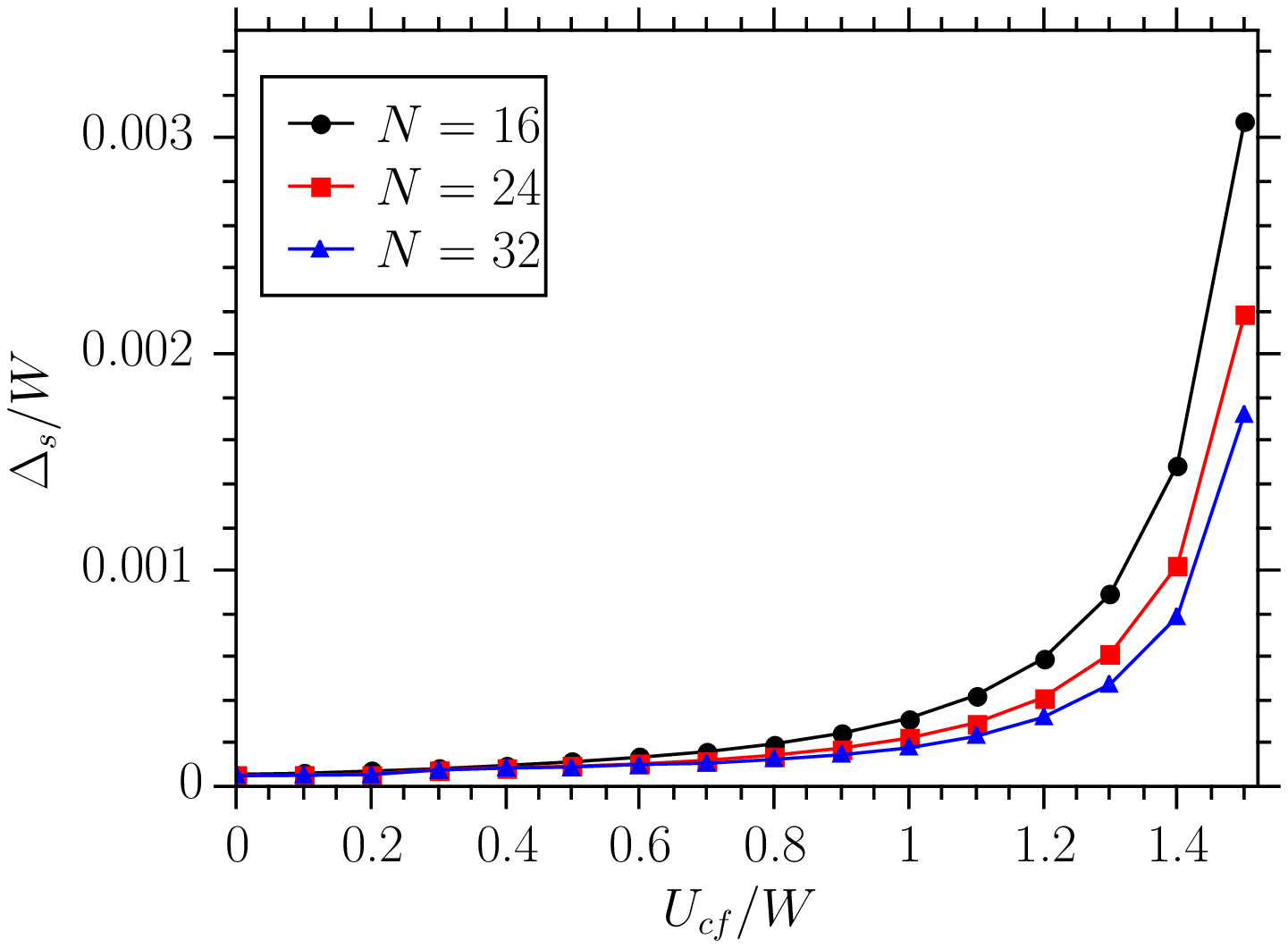}\\
\caption{(color online) The spin gap as a function of $U_{cf}$, for several chainlengths at
$V/W=0.1$, $U_f/W=3$.
}
\label{fig:gap_chain}
\end{figure}

 In the following, we consider the half-filled symmetric EPAM for moderately large
$U_f$.
We address first the effect of $U_{cf}$
on the spin and charge gaps of the EPAM.
The spin gap is defined as
\begin{gather}
 \Delta_s=E_0(S=1,N_{e})-E_0(S=0,N_e),
\end{gather}
where $E_0(S,N_e)$ denotes the ground-state energy in the given $(S,N_e)$ subspace,
where $S$ and $N_e$
are the total spin and particle number, respectively. The latter one is $N_e=2N$ in the half-filled
case. It is known that the
ordinary PAM possesses an
extra SU(2) symmetry\cite{Ueada:exact}, and the
charge gap can be calculated from the expression:
\begin{equation}
\begin{split}
 \Delta_c&=[E_0(S=0,N_e+2)+E_0(S=0,N_e-2)\\
 &-2E_0(S=0,N_e)]/2.
 \end{split}
\end{equation}
This extra
symmetry, however, is no longer present in the EPAM, and therefore we have to apply
the
general definition of the
charge gap\cite{Guerrero:DMRG}:
\begin{gather}
 \Delta_c=E_n(S=0,N_e)-E_0(S=0,N_e)
 \label{eq:charge_gap}
\end{gather}
where $E_n(S=0,N)$ is the energy of the lowest excited state $|n\rangle$, for which
$S=0$ and $\langle n|\sum_q\rho_q|0\rangle\neq0$, where $\rho_q$ is the
$q$ Fourier component of the charge density operator:
\begin{equation}
 \rho_q=\sum_{i=0}^{N-1}e^{-iqr_i}\left[\hat{c}_{i\uparrow}^{\dagger}
	   \hat{c}^{\phantom \dagger}_{i\uparrow}+ \hat{c}_{i\downarrow}^{\dagger}
	   \hat{c}^{\phantom \dagger}_{i\downarrow}+\hat{f}_{i\uparrow}^{\dagger}
	   \hat{f}^{\phantom \dagger}_{i\uparrow}+ \hat{f}_{i\downarrow}^{\dagger}
	   \hat{f}^{\phantom \dagger}_{i\downarrow}\right].
\end{equation}
This definition of the charge gap is motivated by the fact that in optical measurements this gap is
obtained by measuring the conductivity, which is related to the charge density.
The spin and charge gaps as a function of $U_{cf}$ are shown in Fig.
\ref{fig:gaps}.  The charge gap is not calculated far below the crossing point, since the
corresponding
charge excitation lies much higher than the spin gap.
As for the spin gap, it gets very small for small values of $U_{cf}/W$. As shown in Fig.
\ref{fig:gap_scaling} a linear extrapolation in $1/N$ to infinite chain length seems to give a
finite spin gap for any $U_{cf}/W$. 
The number of data points used in the extrapolation depends on
the  scaling of the block entropy and the a priori set accuracy threshold value. 
As the gap gets smaller, it becomes increasingly difficult to determine it accurately for small
values of $U_{cf}$ using the DMRG algorithm. Nevertheless, we see a clear tendency as 
demonstrated in Fig. \ref{fig:gap_chain} that the spin gap increases
monotonically with $U_{cf}/W$ for all system sizes we considered. Since we know that the spin gap
is finite for $U_{cf}=0$, $\Delta_s\sim e^{-\pi tU_f/4\alpha V^2}$, where $\alpha$ is
a scaling constant,\cite{Ueada:exact} we conclude that the spin gap is always finite.
 In the one-dimensional model, in
contrast to
the DMFT results,\cite{Kawakami:Udf_ins}
there is no signature of quantum phase transition and the ground state is always a spin singlet.
\par Since $\Delta_s$ increases with $U_{cf}$, while $\Delta_c$ decreases, they  cross at a value
$U_{cf}^{\rm
cr}\approx U_f/2+W/4$. It can be seen that this crossing point slightly shifts toward smaller
$U_{cf}$ values as $V$ is increased. 
The sharp increase of the spin gap around and above the crossing point
can be understood as follows. Here
 the c and f electrons try not to occupy the same site. Since the total number of f electrons is $N$ in the symmetric
 model, and the same holds for the conduction electrons, the two kinds of electrons can best avoid each other by dominantly
 occupying every second site, the odd sites with two f electrons and the even sites with two
conduction electrons, or vice versa. Since the doubly
 occupied sites are necessarily in a singlet state, a spin flip can be achieved by transferring an
electron to another site and a
 local singlet has to be broken up for that.
For $U_{cf}$ larger than the value at the crossing point, there are a large number of singlet
excited states with lower energy. They determine the density-density correlations and therefore the
gap to the lowest of them is considered as the charge gap, in agreement with Eq.
(\ref{eq:charge_gap}). This gap decreases and tends to zero for $U_{cf}\rightarrow\infty$.

\section{Correlation functions and entanglement patterns}
\subsection{Correlation functions}
\begin{figure}[!htb]
\includegraphics[scale=0.5]{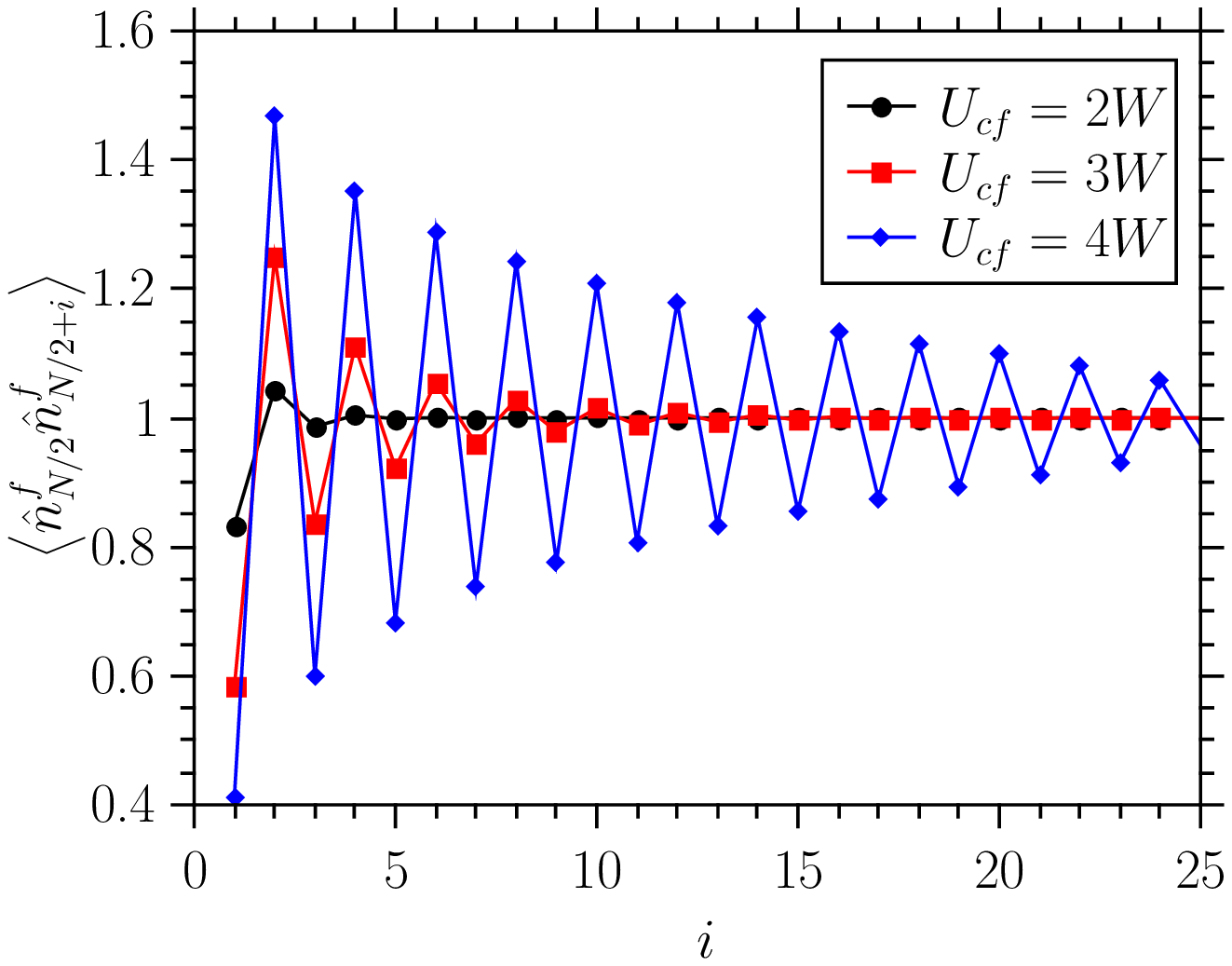}
\caption{(color online) The density correlation
function for different values of $U_{cf}$ and $N=50$ chain length.
The other
parameters are fixed at $V/W=0.1$ and $U_f/W=3$. The lines are
guides to the eye.}
\label{fig:correlations_nolog}
\end{figure}

\begin{figure}[!htb]
\includegraphics[scale=0.5]{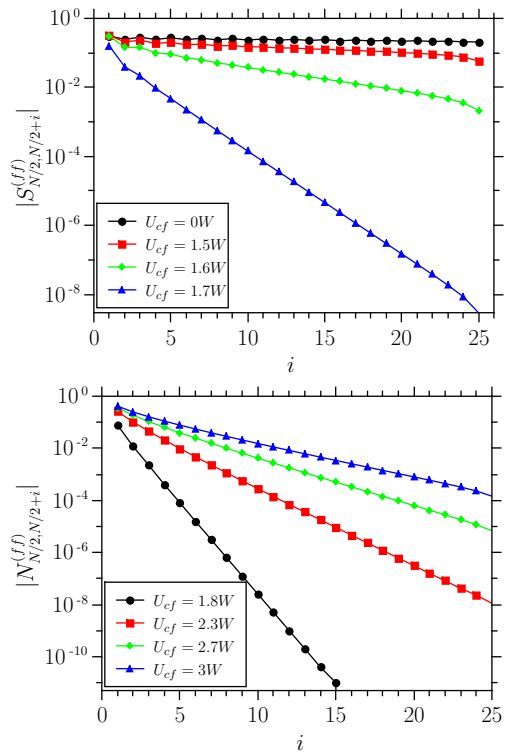}
\caption{(color online) The upper and lower panel show the spin and density correlation
functions respectively for different values of $U_{cf}$ and $N=50$ chain length.
The other
parameters are fixed at $V/W=0.1$ and $U_f/W=3$. The lines are
guides to the eye.}
\label{fig:correlations}
\end{figure}

As a next step we investigate the correlation functions, which provide a further
insight
into the effects of interorbital interaction.
The spin and density
correlation functions are defined in the usual way:
\begin{align}
S^{(ab)}_{ij}&=\left\langle\hat{\boldsymbol{S}}^{(a)}_{i}\hat{\boldsymbol{S}}^{(b)}_j \right\rangle
=\left\langle\frac{1}{2}(a^{\dagger}_{i\uparrow}a_{i\downarrow}b^{\dagger}_{j\downarrow}b_{
j\uparrow}+a^{\dagger}_{i\downarrow}a_{i\uparrow}b^{\dagger}_{j\uparrow}b_{j\downarrow}
)\right.\nonumber\\
&+\left.\frac{1
} {4}(n^{(a)}_{i\uparrow}-n^{(a)}_{i\downarrow})(n^{(b)}_{j\uparrow}-n^{(b)}_{j\downarrow})\right\rangle\\
N^{(ab)}_{ij}&=\left\langle\hat{n}^{(a)}_{i}\hat{n}^{(b)}_j
\right\rangle-\left\langle\hat{n}^{(a)}_{i}\right\rangle\left\langle\hat{n}^{(b)}_{j}\right\rangle,
\end{align}
where $a$ and $b$ stand for either f or c and
$\hat{n}_i^{(a)}=\sum_{\sigma}\hat{n}^{(a)}_{i\sigma}$.
The  correlation functions oscillate with periodicity $2a$ ($a$ is the lattice constant) as seen
in Fig. \ref{fig:correlations_nolog} and their amplitude decays exponentially since the system is
gapped for any value of $U_{cf}$. 
The behavior of the amplitude of the correlation functions is shown in Fig. \ref{fig:correlations}.
The decrease or increase of the gap with $U_{cf}$ is reflected in the variation of the decay length.
 A naive estimate of the correlation length from the linear fit to the data would
give $\xi/a\approx200$, for
$U_{cf}=0$. Although such a large decay length cannot be obtained reliably from calculations on
chains with 50 sites, it can be taken as an order of magnitude estimate, corroborating our
expectations. Namely, such large values of $\xi$ can be explained with the known behavior of the
ordinary PAM. As 
has been pointed out,\cite{Santos:DMRG} the correlation length of the
spin correlation function
increases exponentially by increasing $U_f$.
\par As long as $U_{cf}$ is below the crossing point the spin correlation
function is dominant due
to the tiny spin gap.  The spin correlation
length decreases while the density correlation length
increases as $U_{cf}$ is increased. Above the crossing point, the
density correlation function has the
longer correlation length which corresponds to the peculiar behavior of the
charge gap. 
Knowing that one cannot have true long-range order in one dimension, but
slowly decaying correlations might indicate ordering in higher dimensions, we might guess
that strong $U_{cf}$ leads to charge ordering, as indeed it was found in the DMFT
calculation.\cite{Kawakami:Udf_ins} 
\par
\begin{figure}[!htb]
\includegraphics[scale=0.5]{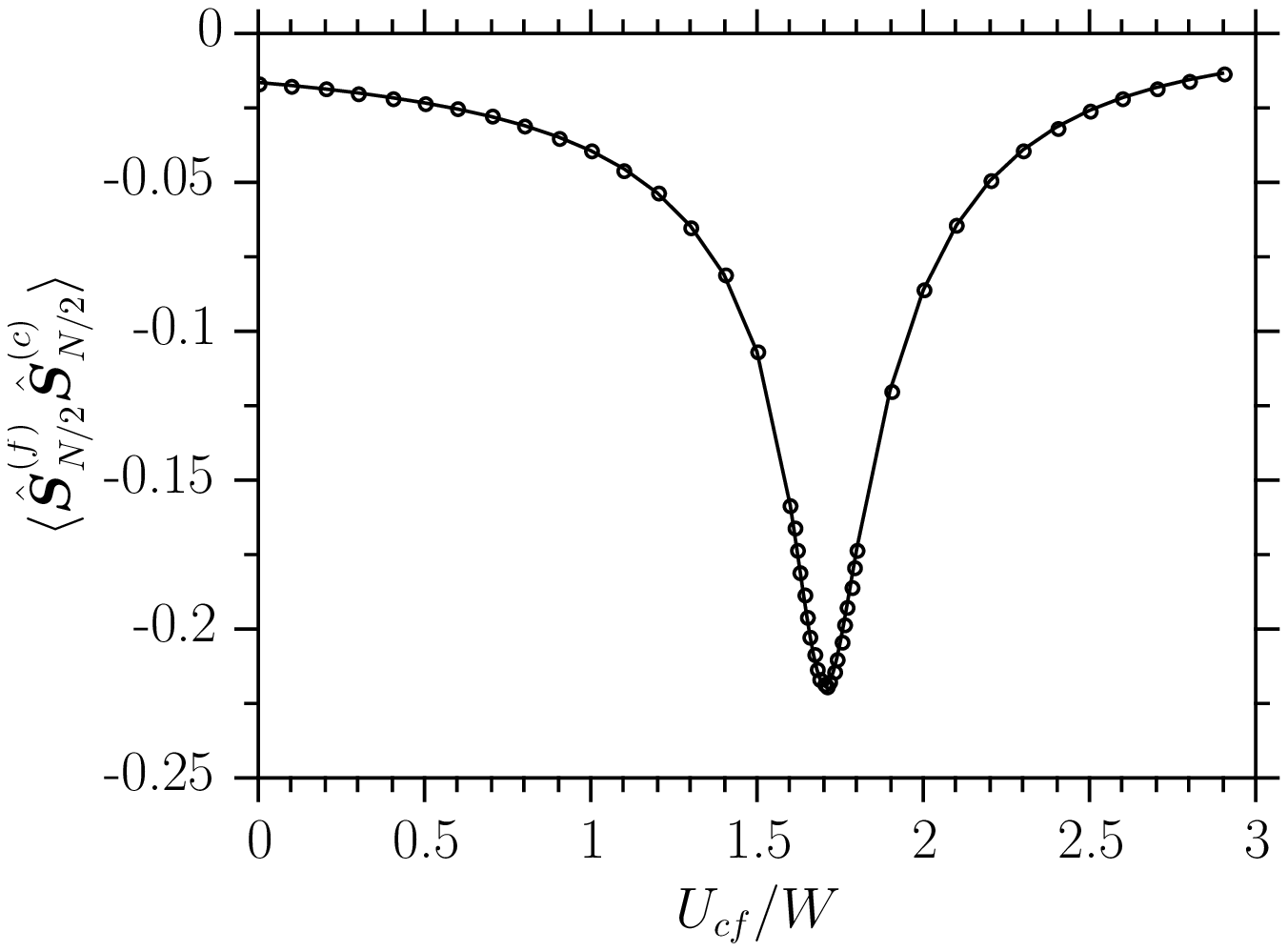}
\caption{The onsite spin correlation between conduction and f electrons (extrapolated to
the thermodynamic limit) for $V=0.1W$ as a
function of $U_{cf}$ and
$U_f/W=3$. The lines are
guides to the eye.}
\label{fig:sfsd}
\end{figure}

 It is interesting to
examine the spin correlation between f and conduction electrons on the same site. This is
shown in Fig. \ref{fig:sfsd}.
It is obvious from the figure that  the local singlet correlation is enhanced
significantly near
$U_{cf}^{\rm cr}$, which confirms the DMFT results.\cite{Kawakami:Udf_ins} 
This behavior is not
expected in the
conventional PAM, since it can be mapped to the Kondo lattice model with weak Kondo
coupling, and the appearance of a Kondo-singlet-like state is expected at strong Kondo couplings.
Therefore
the corresponding Kondo-coupling becomes much stronger at the crossing point. 

\subsection{Quantum information analysis} 
In this section we investigate the behavior of the von Neumann entropies of various subsystem
configurations, which are very reliable tools for detecting drastic changes of the ground state and
analyzing its structure.
We
examined the one-site $s_i$
and two-site $s_{ij}$ entropies, which can be obtained from the appropriate
reduced
density matrices.\cite{legeza2003b,legeza2006} The entropy of a single site can be obtained as
\begin{equation}
 s_i=-{\rm Tr} \rho_i\ln\rho_i,
\end{equation}
 where $\rho_i$ is the reduced density matrix of site $i$,
which is derived from the density matrix of the total system by tracing out the configurations of
all other sites. We also define the c- 
and f-parts of the site entropies ($s^{(c)}_i$, $s^{(f)}_i$) in the following way:
\begin{align}
 s_i^{(c)}=-{\rm Tr} \rho_i^{(c)}\ln\rho_i^{(c)},\\
 s_i^{(f)}=-{\rm Tr} \rho_i^{(f)}\ln\rho_i^{(f)}, 
\end{align}
where $\rho_i^{(c)}$ ($\rho_i^{(f)}$) is obtained by performing an additional trace over the
remaining f (c) degrees of freedom. The two-site entropy is written as
\begin{equation}
 s_{ij}=-{\rm Tr}
\rho_{ij}\ln\rho_{ij},
\end{equation}
where $\rho_{ij}$ is the two-site reduced density matrix of sites $i$ and
$j$. We can also introduce the partial two-site entropies for c or f electrons on site $i$
and 
c or f electrons on site $j$:
\begin{equation}
 s_{ij}^{(ab)}=-{\rm Tr}
\rho_{ij}^{(ab)}\ln\rho_{ij}^{(ab)}, \quad a,b\in\{c,f\}
\end{equation}
where $\rho_{ij}^{(ab)}$ is derived from $\rho_{ij}$ by tracing out the states of the other electrons. The DMRG implementation described in the beginning of Sec. II.
enables us to determine $s^{(a)}_i$ and $s_{ij}^{(ab)}$  separately.
The mutual information which measures the entanglement between sites $i$ and $j$ can be obtained from:
\begin{gather}
   I_{ij}=s_i+s_j-s_{ij},
\end{gather}
while the mutual
information between $a$ and $b$ type electrons on sites $i$ and $j$ is defined as
\begin{gather}
   I_{ij}^{(ab)}=s_i^{(a)}+s_j^{(b)}-s_{ij}^{(ab)},
\end{gather}
which measures the entanglement between $a$ and $b$ type electrons on sites $i$ and $j$. 
\par
\begin{figure}[!htb]
\includegraphics[scale=0.5]{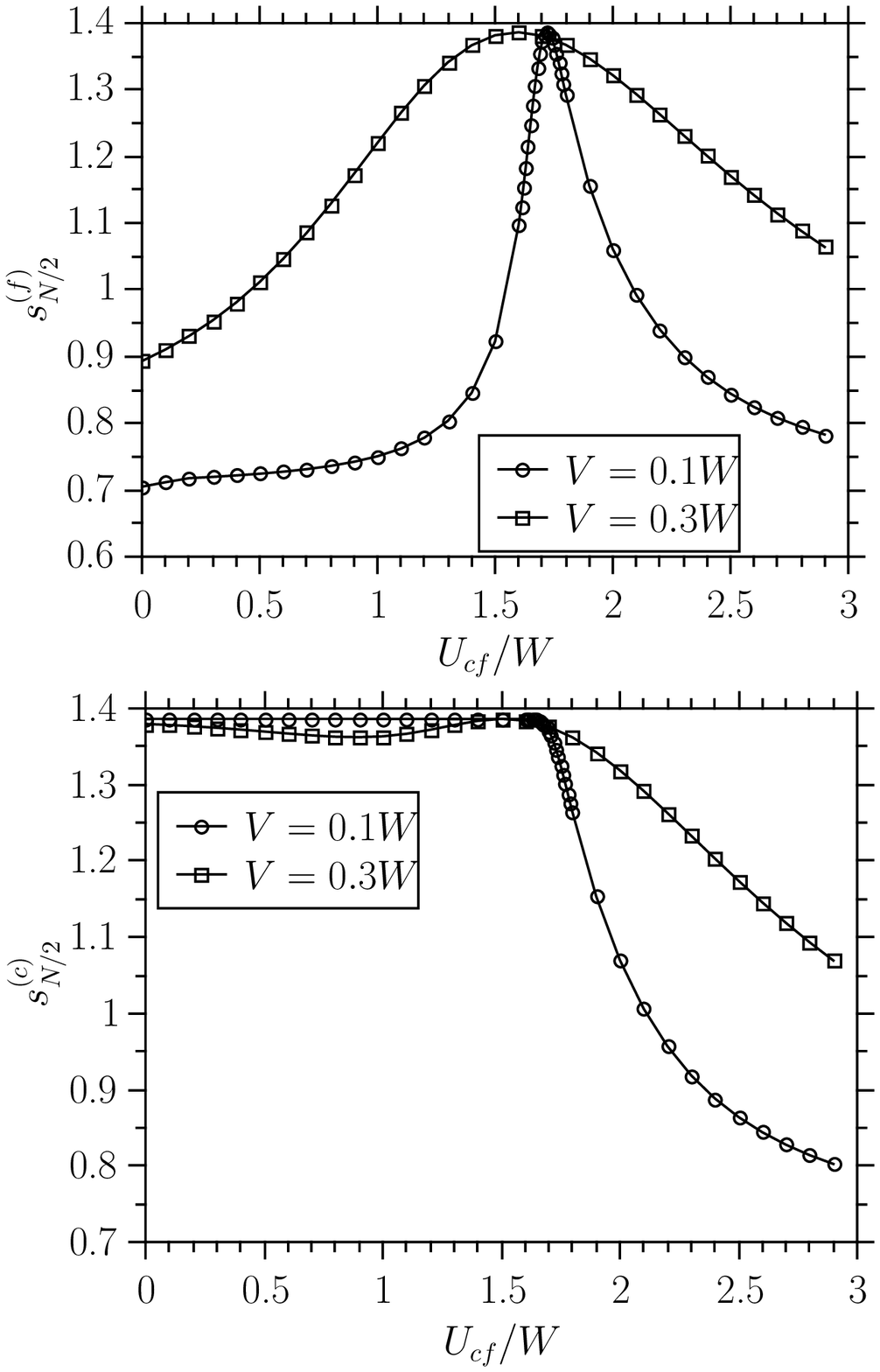}
\caption{The extrapolated f (upper panel) and c (lower panel) site entropies as a function
of $U_{cf}$, for
$U_f/W=3$. The lines are
guides to the eye.}
\label{fig:entropy}
\end{figure}

 At first we
consider
the single site entropies, which are
shown in Fig. \ref{fig:entropy}.
For small $V$ the entropy of f electrons starts from slightly above $\ln 2$. When $U_{cf}=0$ the f electrons
are
strongly
correlated and since the f-level occupancy must be exactly one (in the symmetric model), only one
electron
with up or down spin can
occupy the f level. Due to the small c-f hybridization the entropy of f electrons is slightly
higher than
$\ln2$, since a small number of doubly occupied levels can also be present. Switching on
$U_{cf}$
leads to the appearance of more and more doubly occupied f sites, so the entropy begins to
increase. At a certain
value of $U_{cf}$, which is the crossing point defined earlier, a peak is developed,
where the
entropy of f electrons takes its maximum value, $\ln4$, then it begins to decrease and approaches $\ln2$
again, since for large $U_{cf}$ an f site is expected to be either doubly occupied or
empty. For larger hybridization this
sharp maximum is significantly broadened. Concerning the conduction electrons, their entropy is
$\ln4$ as long as
$U_{cf}<U_{cf}^{\rm cr}$, since they are free particles. Above the crossover value, the
probability of finding zero or two c electrons on the same site  increases from 1/4 to 1/2, so correlation is
developed between c electrons. 
It is readily observed that there should be a remarkable
change in the ground state, where the entropy of f electrons has a maximum. 
\par
\begin{figure}[!htb]
\includegraphics[scale=0.5]{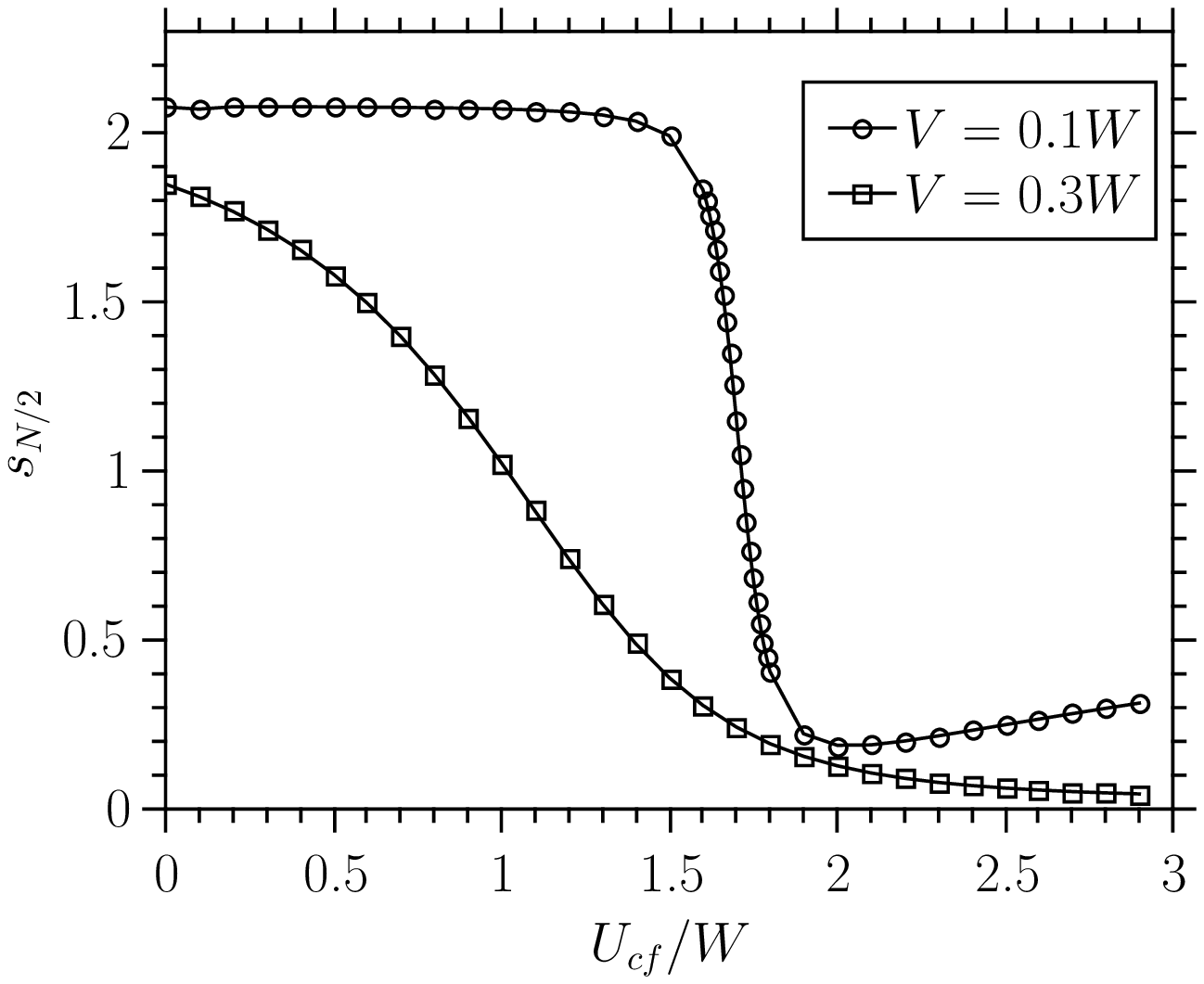}
\caption{The extrapolated one-site entropies, for two different values
of hybridization and
$U_f/W=3$. The lines are
guides to the eye.}
\label{fig:two_site_entropy}
\end{figure}

 We also examined the one-site entropy of the EPAM,
and it is shown in Fig. \ref{fig:two_site_entropy}.
For weak hybridization it drops rather drastically around $U_{cf}$ corresponding to the crossover
point and then starts to increase slowly, while for stronger hybridization the decrease of the site
entropy
is less drastic. But in both cases $s_i$ is much smaller above $U_{cf}^{\rm cr}$ than below this
value. 
A nearly vanishing $s_i$ is indication that the wave function is dominated by terms localized to
this site.
\par
\begin{figure}[htb]
\centerline{
\scalebox{0.7}{\includegraphics{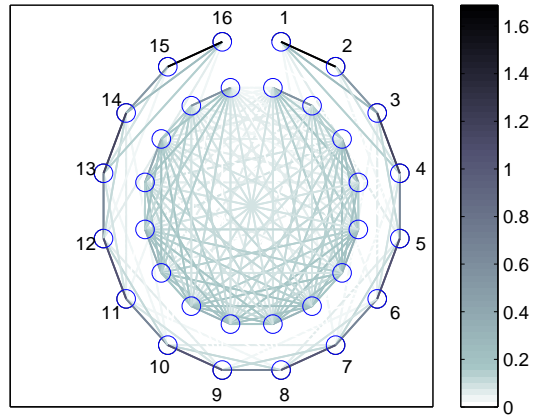}}
} 
\caption{(color online) Schematic view of all components of the mutual information ($I_{ij}^{(cc)}$,
$I_{ij}^{(cf)}$, $I_{ij}^{(ff)}$) for $U_{cf}=0$,
$V/W=0.1$, 
$U_f/W=3$ and $N=16$. The inner and outer circles denote f and c sites, respectively. The numbers
denote the real EPAM sites.}
\label{fig:mutual_inf_Ucf0}
\end{figure}

 To better reveal the reason of these anomalies we calculated the mutual information
between c and f type
electrons on sites $i$ and $j$ for several
values of $U_{cf}$. The mutual information for $U_{cf}=0$ is shown in Fig.
\ref{fig:mutual_inf_Ucf0} for weak hybridization.
It is easy to observe that moderately strong but short-ranged entanglement is developed between c
electrons and much weaker between c and f electrons due to the small hybridization. On the other
hand long-ranged but weaker entanglement is formed between the f electrons. This is the consequence
of the strong
RKKY-interaction, which results in the antiferromagnetic correlations between the f
electrons; the f electrons form a collective singlet. 
\par
\begin{figure}[!htb]
\centerline{
\scalebox{0.7}{\includegraphics{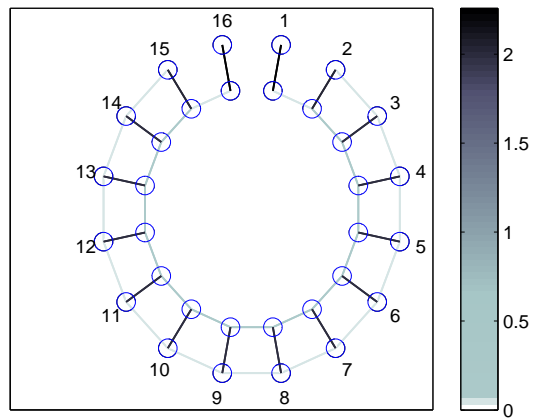}}
} 
\caption{(color online) Schematic view of all components of the mutual information ($I_{ij}^{(cc)}$,
$I_{ij}^{(cf)}$, $I_{ij}^{(ff)}$) for $U_{cf}=1.75W$,
$V/W=0.1$,
$U_f/W=3$ and $N=16$. The inner and outer circles denote f and c sites respectively.  The numbers
denote the real EPAM sites.}
\label{fig:mutual_inf_Ucf1p75}
\end{figure}

\begin{figure}[!htb]
\centerline{
\scalebox{0.7}{\includegraphics{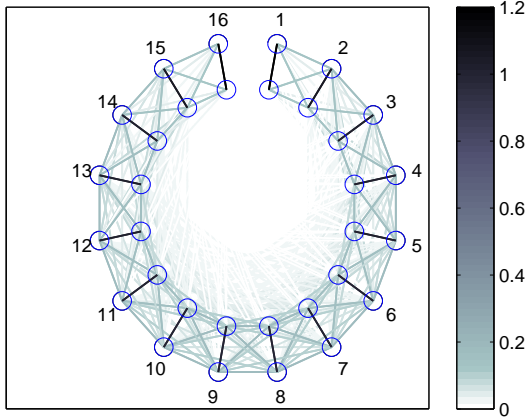}}
} 
\caption{(color online) Schematic view of all components of the mutual information ($I_{ij}^{(cc)}$,
$I_{ij}^{(cf)}$, $I_{ij}^{(ff)}$) for $U_{cf}=4W$,
$V/W=0.1$,
$U_f/W=3$ and $N=16$. The inner and outer circles denote f and c sites respectively.  The numbers
denote the real EPAM sites.}
\label{fig:mutual_inf_Ucf4}
\end{figure}

 The mutual information diagram
has an entirely different structure for $U_{cf}=1.75W\approx U_{cf}^{\rm cr}$, which is
shown in Fig. \ref{fig:mutual_inf_Ucf1p75}.
One can see that there is a strong entanglement between  c and f electrons on the same site. The
 entanglement bonds between the sites are much
weaker.
According to
the
entanglement map, the ground-state wave function becomes approximately a product state.   We will
determine the structure of the onsite state later. Lastly, we consider the case when
$U_{cf}=4W>U_{cf}^{\rm cr}$. The
mutual information is shown in Fig. \ref{fig:mutual_inf_Ucf4}.
Here we observe that the strong onsite entanglement remains, however, moderately strong
bonds appear between neighbouring c and f sites. It is worth noting, that the magnitude of the
entanglement of the onsite bonds is
$\mathcal{O}(1)$, while it is $\mathcal{O}(10^{-3})$ for the next largest entanglement bond when
$U_{cf}=3W$, $V=0.3W$ and $U_f=3W$, that is, every other bond is smaller by two orders
of magnitude.
\par The above statements can be quantified if we introduce the following quantitiy:
\begin{align}
 I_{\rm dist}^{(m)}=&\frac{1}{N}\sum_{ab}\sum_{ij}I^{(ab)}_{ij}(i-j)^{m}.
\end{align}
$I_{\rm dist}^{(m)}$ is the $m$th momentum of the distribution $I^{(ab)}_{ij}$, which measures the
localization of the entanglement. Its
values for different values of $U_{cf}$ and $m$ are shown in Table  \ref{table:I}.
\begin{table}[h]
\begin{tabular}{@{}c@{\hspace{4mm}}c@{\hspace{4mm}}c@{\hspace{4mm}}}
\toprule
$U_{cf}/W$ &   $I_{\rm dist}^{(2)}$  & $I_{\rm
dist}^{(4)}$\\ \hline
$0$ &  $50.7$ &  $4169.9$\\
$1.75$ &  $0.24$ &  $0.38$ \\
$4$ &  $19.1$ & $959.0$\\
\toprule
\end{tabular}
\caption{The different momenta of $I_{ij}^{(ab)}$ for several values of $U_{cf}$ and $V=0.1W$.}
\label{table:I}
\end{table}
It is easily seen that the entanglement is most localized when $U_{cf}=1.75W$. In
the other two cases the entanglement is much more delocalized. 
\par
\begin{figure}[!htb]
\includegraphics[scale=0.5]{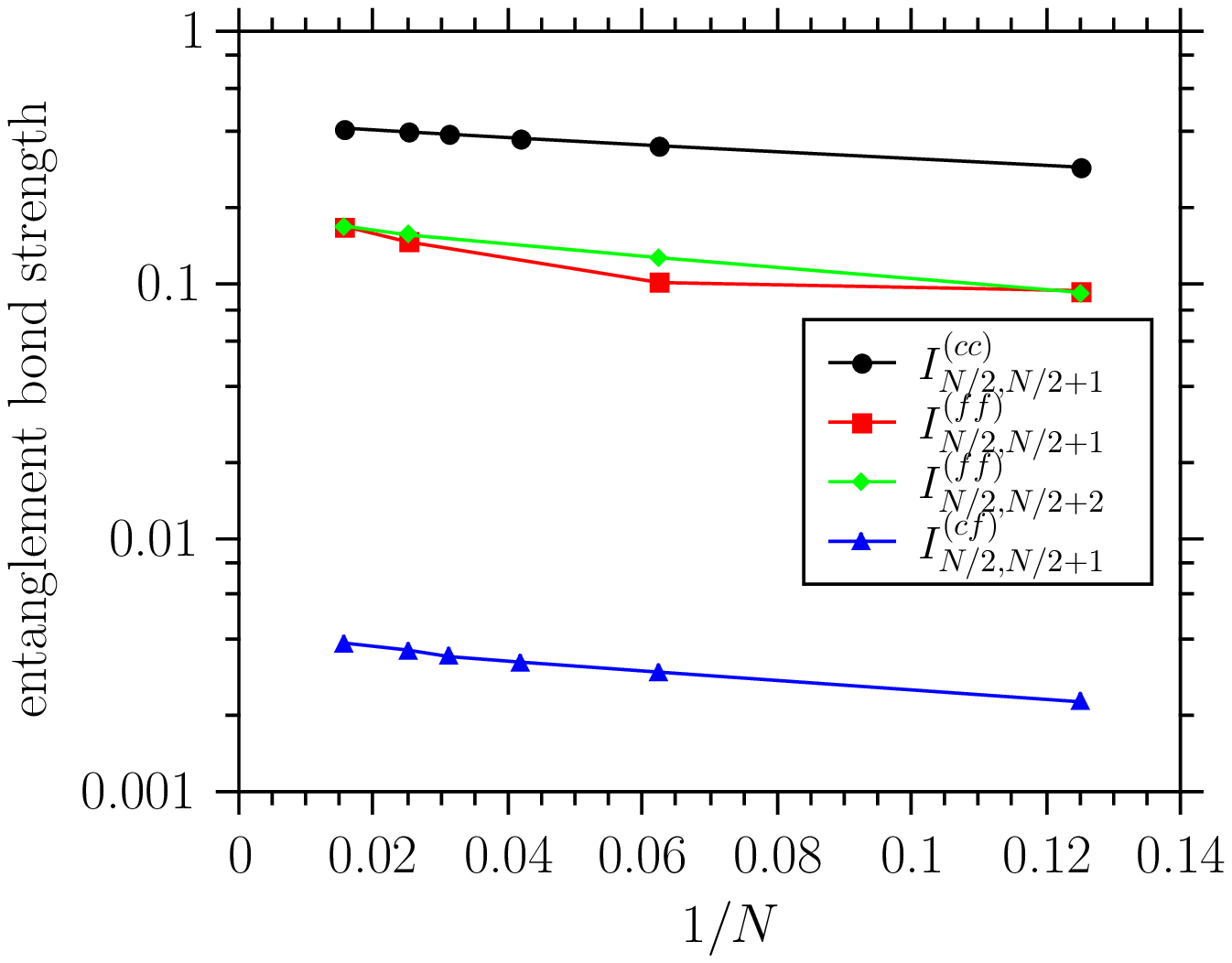}
\caption{(color online) The finite-size scaling of various entanglement bond for $U_{cf}=0$, $V=0.1W$ and
$U_f/W=3$. The lines are
guides to the eye.}
\label{fig:bond_scaling}
\end{figure}

 These conclusions have been obtained for finite
systems, therefore we have to investigate the finite-size effects. For $U_{cf}=0$ the entanglement
bonds show strong dependence on the chain length, while for $U_{cf}=1.75W$ and $4W$ the above
results are very close the bulk limit for $N=16$ already. The size dependence of the bonds for
$U_{cf}=0$ is shown in Fig. \ref{fig:bond_scaling}.
\par One can naturally ask what the relevant physical process is in creating the strong onsite bonds
around the crossing point and for stronger $U_{cf}$. To answer
this question we examined the eigenvalues ($\omega_{\alpha}$, $\alpha=1,\dots,16$) of the two-site
density matrix $\rho_{ij}^{(ab)}$. For $U_{cf}=0$ we found that several eigenvalues
of $\rho_{N/2N/2}^{(cf)}$ are of the same order of magnitude and the eigenvectors contain all basis
states $|\alpha_c,\alpha_f\rangle$, except for $|0,0\rangle$ and
$|\uparrow\downarrow,\uparrow\downarrow\rangle$.
However, for $U_{cf}=1.75W$ one of the eigenvalues of
$\rho_{N/2N/2}^{(cf)}$ becomes almost two orders of magnitude larger than the others. The
corresponding eigenfunction reads:
\begin{equation}
\label{eq:wavefunction:Ucf1.75}
\begin{split}
 \phi^{(cf)}_{N/2N/2}=&-0.5798(|\uparrow\downarrow,0\rangle+|0,
\uparrow\downarrow\rangle)\\
&-0.4048(|\uparrow,\downarrow\rangle-|\downarrow,\uparrow\rangle).
\end{split}
\end{equation}
It is seen that the configuration in which the site is occupied with two f electrons in a singlet
state or two conduction electrons in a singlet state has roughly the same weight as the state in
which a conduction and an f electron form a singlet. The latter configuration was
observed in the spin correlation function in Fig. \ref{fig:sfsd} and corresponds to the enhanced
spin Kondo-effect. The former one describes the fluctuations between the doubly occupied and empty
c-f configurations.
Further increase of $U_{cf}$ results in the suppression of the singlet part in Eq.
(\ref{eq:wavefunction:Ucf1.75}). For example for $U_{cf}=4W$, the eigenfunction corresponding to the
most significant
eigenvalue is:
\begin{equation}
\label{eq:wavefunction:Ucf4}
\begin{split}
 \phi^{(cf)}_{N/2N/2}=& \ 0.7049(|\uparrow\downarrow,0\rangle+|0,
\uparrow\downarrow\rangle)\\
&-0.0561(|\uparrow,\downarrow\rangle-|\downarrow,\uparrow\rangle).
\end{split}
\end{equation}
Parallel with the suppression of the c-f singlets, two f or
two conduction electrons occupy more and  more this site. This is the reason why strong onsite
entanglement bonds appear in
Fig. \ref{fig:mutual_inf_Ucf4}. 
\par
\begin{figure}
 \scalebox{0.9}{\includegraphics[scale=0.45]{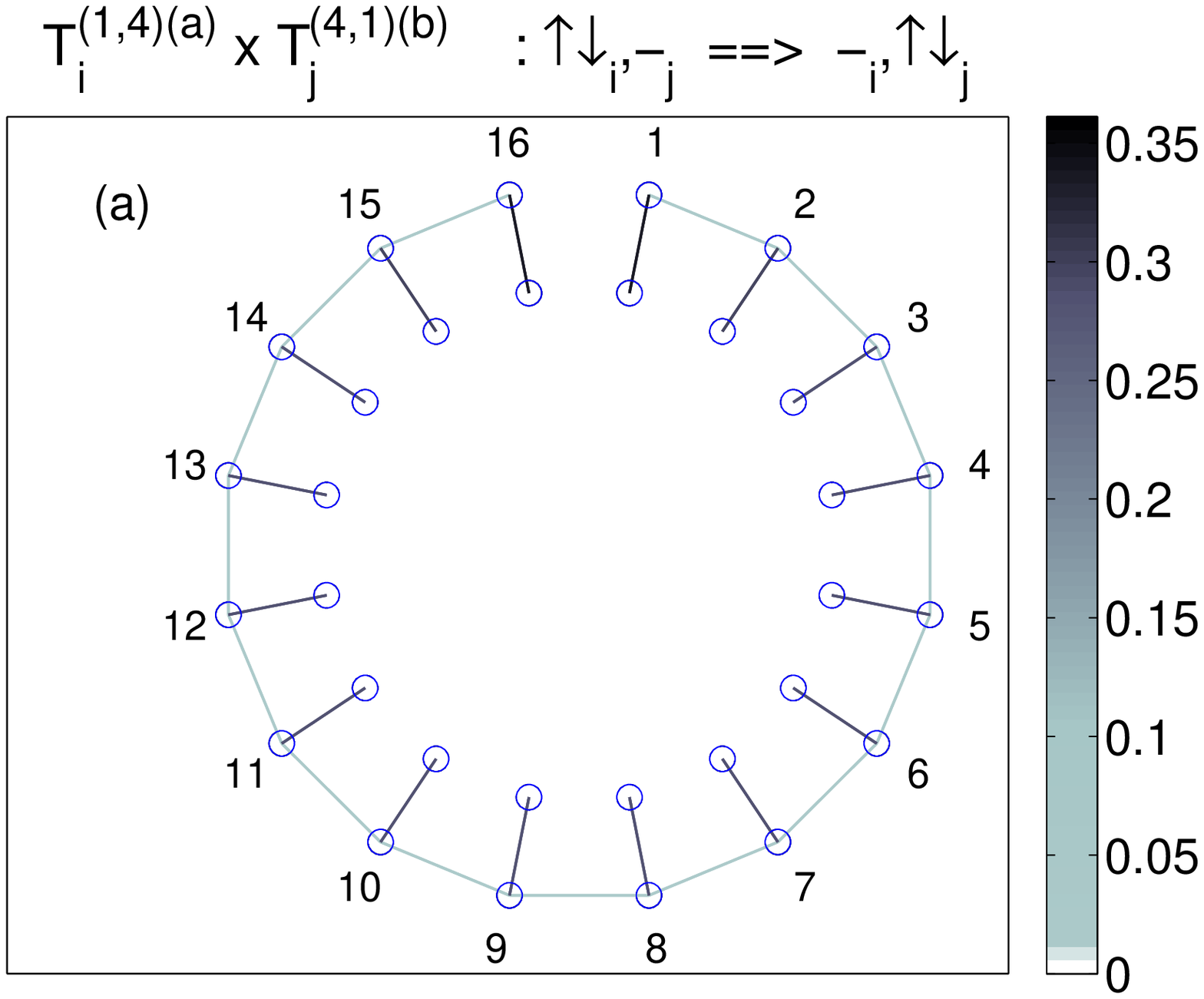}}\\
 \scalebox{0.9}{\includegraphics[scale=0.45]{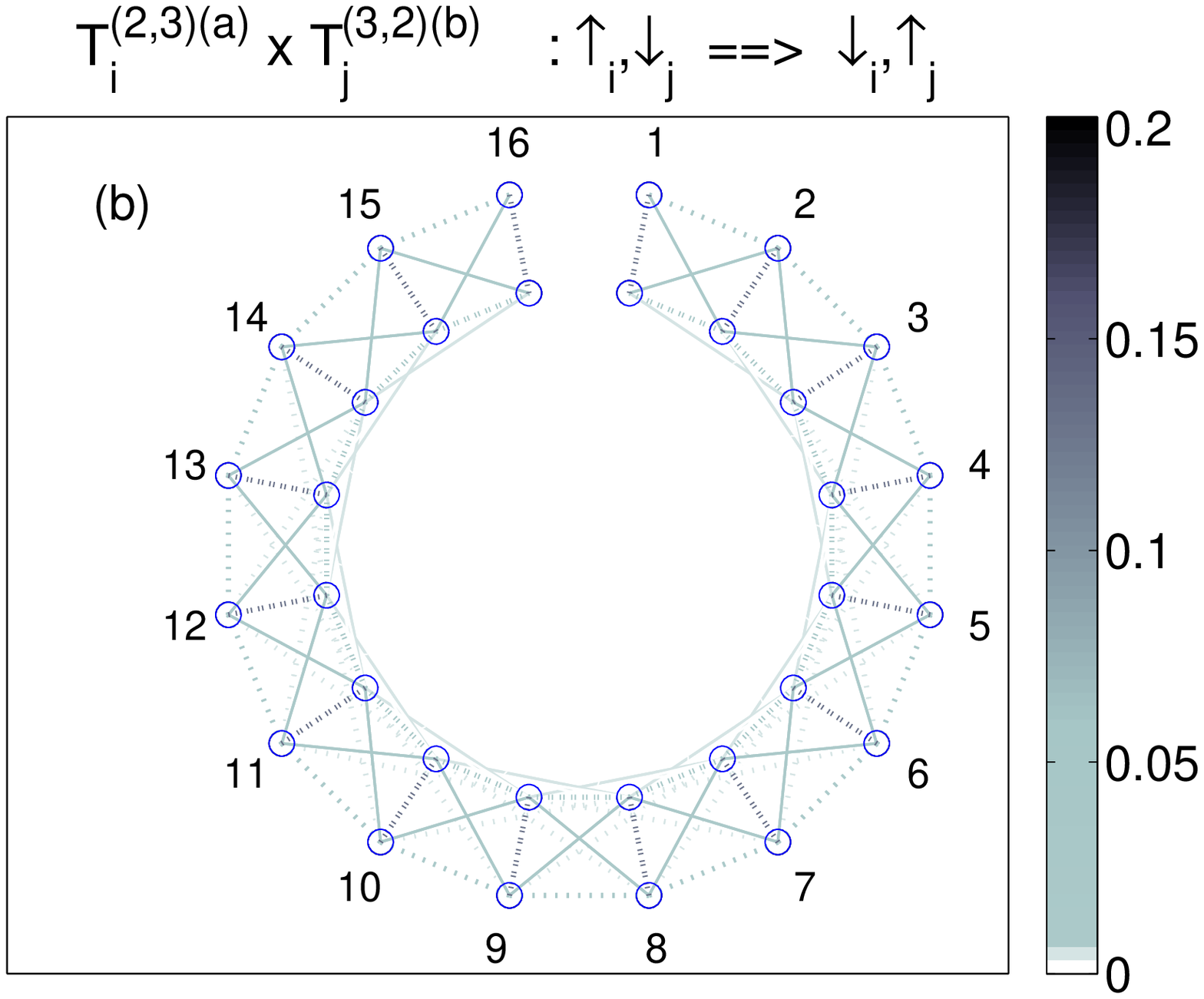}}
   \caption{(color online) Pictorial representation of two generalized correlation functions:
$\langle\mathcal{T}^{(1,4)(a)}_i\mathcal{T}^{(4,1)(b)}_j\rangle_{\rm C}$ (panel (a)) and
$\langle\mathcal{T}^{(2,3)(a)}_i\mathcal{T}^{(3,2)(b)}_j\rangle_{\rm C}$ (panel (b)).
The parameters are $U_{cf}=1.75W$,
 $V/W=0.1$, $U_f/W=3$ and $N=16$. The inner and outer circles denote f and c sites, respectively.
The positive and negative values of the correlations are shown by solid and dotted lines,
respectively. The numbers
 denote the real EPAM sites.}
 \label{fig:gen_corr_fun1}
\end{figure}

\begin{figure}
 \scalebox{0.9}{\includegraphics[scale=0.45]{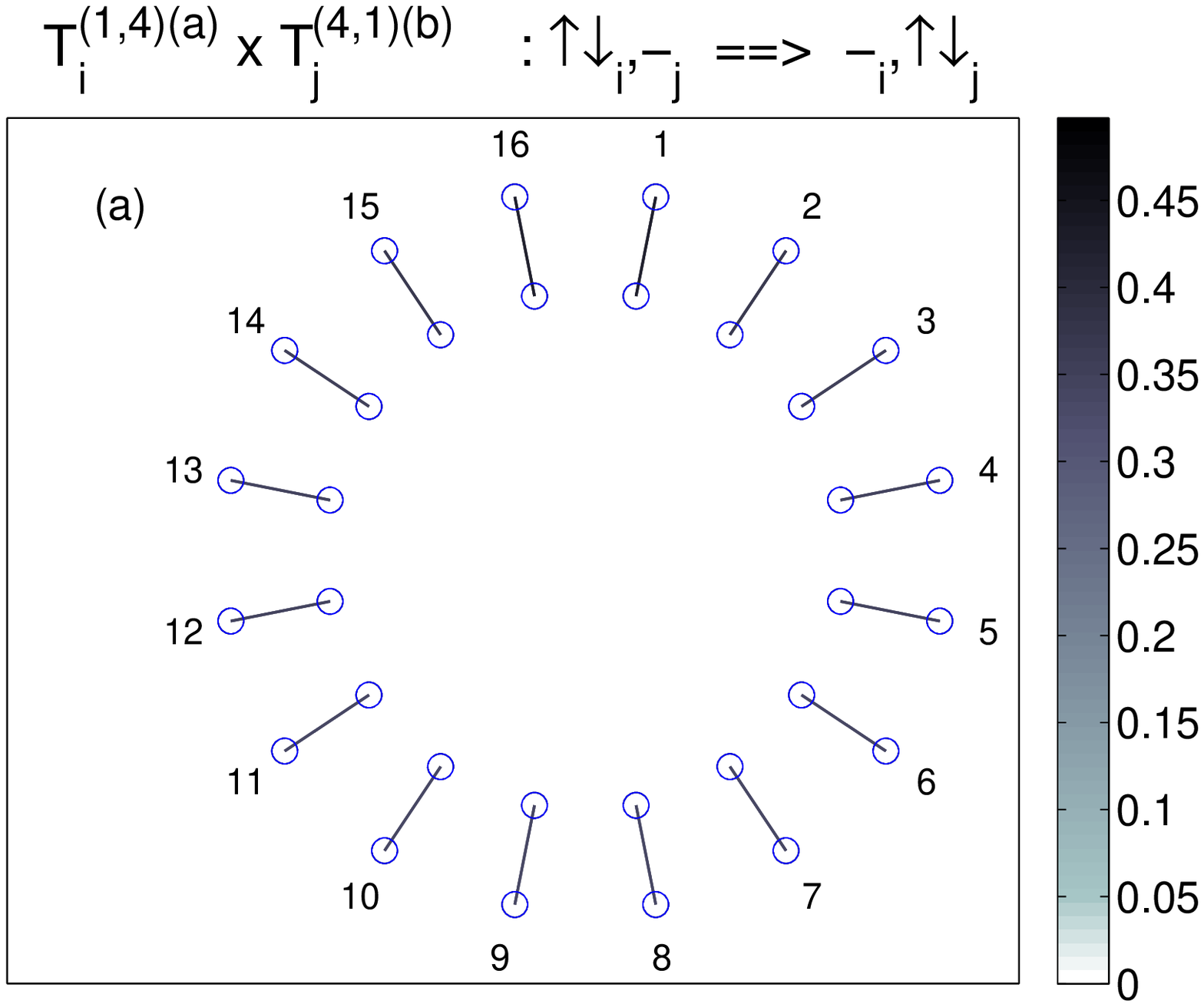}}\\
 \scalebox{0.9}{\includegraphics[scale=0.45]{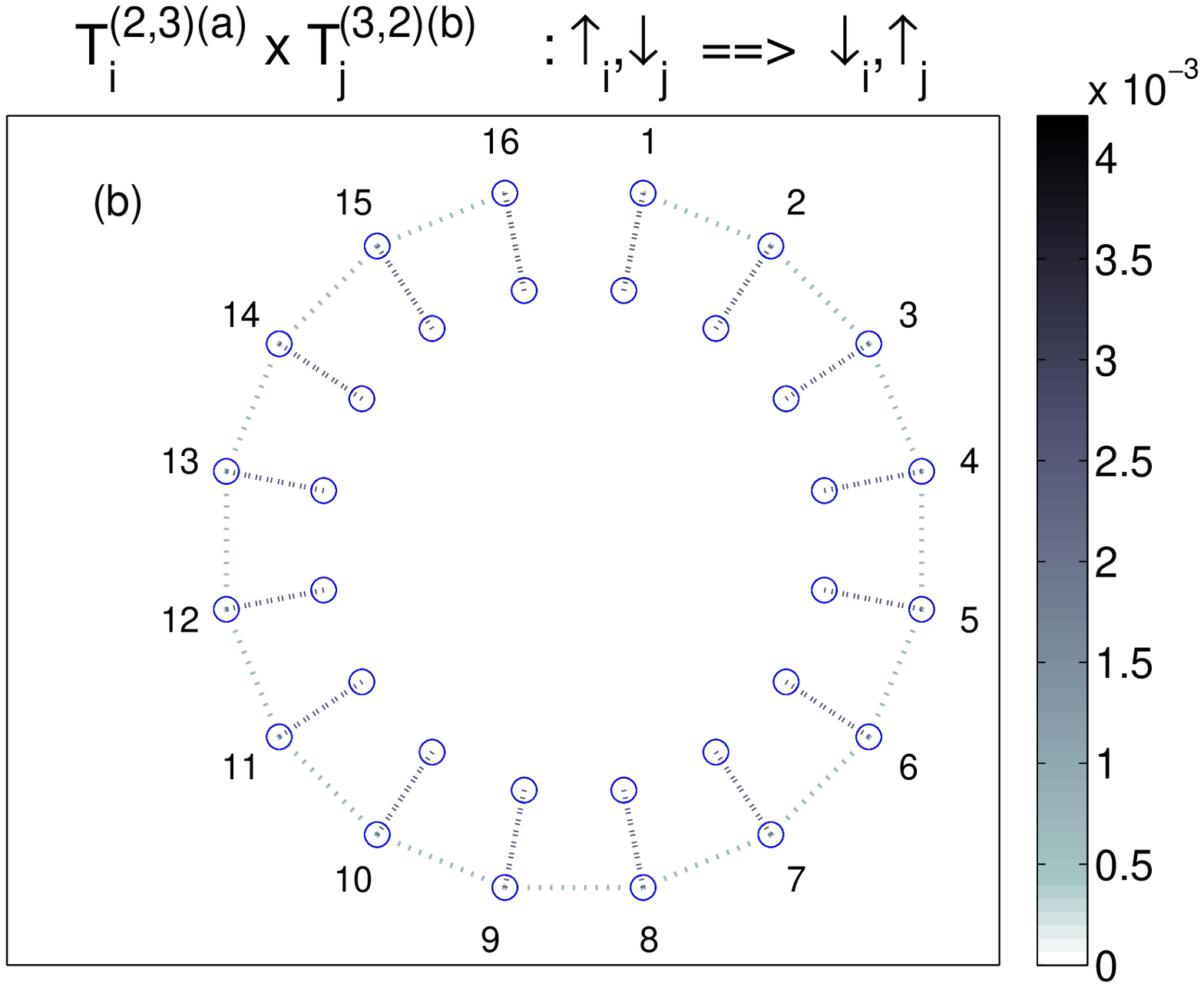}}
   \caption{(color online) The same as in Fig. \ref{fig:gen_corr_fun1}, but for $U_{cf}=4W$.}
 \label{fig:gen_corr_fun2}
\end{figure}

 In addition, as has been shown in Ref.
[\onlinecite{Legeza:entanglement}] one can also analyze the
sources of entanglement encoded in $I_{ij}^{(ab)}$ by studying the behavior of the matrix elements
of $\rho_{ij}^{(ab)}$. They can be expressed in terms of the generalized correlation functions of 
$\mathcal{T}^{\alpha^{\prime}\alpha(a)}_i$ and  $\mathcal{T}^{\beta^{\prime}\beta(b)}_j$,
where $\mathcal{T}^{\alpha^{\prime}\alpha(a)}_i$ is the transition matrix that transfers state
$|\alpha\rangle^{(a)}$ 
($\alpha=1\dots4$ and $a\in\{c,f\}$) defined in Eq. (\ref{eq:states}) into state 
$|\alpha^{\prime}\rangle^{(a)}$ on the same site $i$, while all other matrix elements vanish.
For example 
\begin{equation}
\mathcal{T}_i^{(2,3)(c)}|3\rangle^{(c)}=\mathcal{T}_i^{(2,3)(c)}\hat{c}_{i\uparrow}^{\dagger}
|0\rangle=\hat{c}_{i\downarrow}^{\dagger}|0\rangle=|2\rangle^{(c)}
\end{equation}
and
\begin{equation}
\mathcal{T}_i^{(3,4)(f)}|4\rangle^{(f)}=\mathcal{T}_i^{(3,4)(f)}\hat{f}_{i\uparrow}^{\dagger}\hat{f}
_{i\downarrow}^{\dagger}|0\rangle=\hat{f}_{i\uparrow}^{\dagger}|0\rangle=|3\rangle^{(f)}.
\end{equation}
 We studied the connected
part of the generalized correlation functions, $\langle\mathcal{T}^{\alpha^{\prime}\alpha(a)}_i\mathcal{T}^{\beta^{\prime}\beta(b)}_j
\rangle_{\rm C}=\langle\mathcal{T}^{\alpha^{\prime}\alpha(a)}_i\mathcal{T}^{\beta^{\prime}\beta(b)}_j
\rangle-\langle\mathcal{T}^{\alpha^{\prime}\alpha(a)}_i\rangle\langle\mathcal{T}^{\beta^{\prime}\beta(b)}_j
\rangle$, where the disconnected part, given by the product of the expectation values of local
transition operators, is substracted. We demonstrate here, that these can be used to identify the
relevant physical processes that lead to the generation of entanglement. Namely, we show the
correlation functions of two different transition operators giving the largest contribution to
$\rho_{ij}^{(ab)}$. One of them describes the hopping of a down- and up-spin electron pair, the
other one is a spin-flip. The matrix elements of these operators are shown in Figs.
\ref{fig:gen_corr_fun1} and \ref{fig:gen_corr_fun2}.
It is clearly seen that the correlations in spin-flip processes are significantly
reduced
as $U_{cf}$ is increased. At the same time the fluctuations of the up- and down-spin electron pairs
is enhanced. These results support our previous findings derived from the analysis of the
eigensystem of the two-site density matrix $\rho_{ij}^{(ab)}$.
\par The decay of the mutual information is governed by the smallest gap in
the model. The fastest decay is found for $U_{cf}\approx U_{cf}^{\rm cr}$, where
both the spin and charge gaps are large. For other $U_{cf}$ values we observed a slower decay,
which is in agreement with the behavior of the gaps. We observed that the mutual information decays as the
square of the most slowly decaying correlation functions at long distances in agreement with the results of Ref. [\onlinecite{Legeza:entanglement}].

\section{Conclusions}
In this paper we investigated an extended periodic Anderson model, where the interaction
between conduction and f electrons, $U_{cf}$, has been included. Our aim was to examine the
properties of the model in one dimension by applying the density matrix renormalization
group algorithm. As a first step, we investigated the spin and charge excitations of the
model. It turned out that the model is always gapped. The spin gap, $\Delta_s$, is much smaller than
the charge gap, $\Delta_c$, for small $U_{cf}$. Since $\Delta_s$ increases with $U_{cf}$ while
$\Delta_c$ decreases, they cross at some value of $U_{cf}$ and the charge gap gets smaller. This
result
may give a possible explanation
for the anomalous behavior of the
gaps observed in \mbox{CeRhAs}, where $\Delta_s/\Delta_c>1$ was
measured. The crossing point shifts only slightly as the hybridization becomes stronger.
\par As a next step, the spin and density correlation functions have been determined. Below
the crossing point the spin correlation function is dominant due to strong
antiferromagnetic coupling mediated by the RKKY interaction. As $U_{cf}$ is increased
the spin correlation length decreases, while the density correlation length increases and
becomes dominant above the crossing point. In higher dimensional systems, especially in DMFT
calculations, the dominance of the spin or charge excitations leads to two distinct ordered phases.
In the DMFT results, an antiferromagnetic--charge order transition was found, moreover the
position of this critical point strongly depends on the hybridization. In one dimension we
have found, however, no sign of phase transition. In our model the correlation functions
always decay exponentially, but we can distinguish two regimes depending on which decay length is
larger.
\par Finally, we performed a quantum information analysis to reveal the structural
changes in the ground state wave function. The one-site entropy of c and f electrons varies rapidly around the point 
where the spin and charge gaps are equal. The entropy of f electrons has a maximum there, while the entropy of c electrons
starts to decrease rapidly. We calculated the mutual
information for several c-f
interaction strengths, which measures the entanglement between different sites. It turned
out that for small hybridization the wave function is approximately a product state consisting of
strongly localized states at the sites. For larger $U_{cf}$ the strong onsite entanglement remains,
which originates from the tendency of charge ordering, the preference of having two f electrons on
even sites
and two c electrons on odd sites or vice versa.

\acknowledgments{This work was supported in part by the
Hungarian Research Fund (OTKA) through Grant Nos.~K 100908 and NN110360. The research of I. H. was
supported by
the European Union and the State of Hungary, co-financed by
the European Social Fund in the framework of T\'AMOP-4.2.4.A/ 2-11/1-2012-0001  'National
Excellence Program'. I. H. acknowledges fruitful discussions with A. Kiss and K. Itai.}

\end{document}